\documentclass[mathleft
]{an}
\usepackage{graphicx}
\usepackage{times}
\usepackage{tipa}
\begin{document}

% The following seven commands are intended for editorial usage and should be ignored by
% the author(s).
\Pagespan{789}{}% Document's page range. 
% If second parameter is left empty, the last page is computed automatically.
\Yearpublication{2012}%
\Yearsubmission{2012}%
\Month{99}%   
\Volume{999}%  
\Issue{99}% 
% \DOI{This.is/not.aDOI}% 

\title{No evidence for an early seventeenth-century Indian sighting of\\ Kepler's supernova (SN1604)}

\author{Robert H.\ van Gent\thanks{Corresponding author:
  \email{r.h.vangent@uu.nl}\newline}
%Example 
%for footnote, note the usage of the \texttt{fnmsep}
%command as separator between institute number and footnote mark} 
}
\titlerunning{No evidence for Indian sighting of SN1604}
\authorrunning{R.H.\ van Gent}
\institute{Institute for History and Foundations of Science, Utrecht University, P.O.~Box 80.010, NL-3508~TA Utrecht,\\
           The Netherlands}

\received{2012 May 8}
\accepted{2012 ???~??}
\publonline{later}


\abstract{%
In a recent paper in this journal Sule et al.\ (2011) argued that an early 17th-century Indian mural of the constellation Sagittarius with a dragon-headed tail indicated that the bright supernova of 1604 was also sighted by Indian astronomers. In this paper it will be shown that this identification is based on a misunderstanding of traditional Islamic astrological iconography and that the claim that the mural represents an early 17th-century Indian sighting of the supernova of 1604 has to be rejected.} 

\maketitle

\section{Introduction}

Ancient records from various civilisations frequently note the occurrence of a `new star' (comet, nova or supernova) in the heavens. Examples of such observations can be found in Babylonian, Greek/Roman, Far Eastern (China, Korea \& Japan), Islamic and European sources. Of these cosmic events, the appearance of a supernova (the catastrophic collapse of the core of a massive star at the end of its evolution) is the rarest and the number of reliably observed stellar outbursts is small. 

As all observed supernovae in our own galaxy (with the possible exception of the progenitor star of the radio/X-ray source Cas~A, which may have been observed in August 1680 by John Flamsteed) occurred before the invention of the telescope (in 1608), each newly discovered record of such an event can be very useful for modern astronomers.

The last observed supernova in our own galaxy was first seen by two Italian astronomers on 9~October 1604 (New Style) in the constellation Ophiuchus, then also commonly known as Serpentarius (``The Snake-Bearer''). Now known as SN1604, V843~Oph or Kepler's supernova, the new star attracted much attention among European astronomers and astrologers as it occurred in close proximity with the planets Mars, Jupiter and Saturn and some (among whom was Johannes Kepler) believed that the object was actually the result of this very close planetary configuration.

After its discovery, it continued to brighten and reached a peak magnitude of about $-3.0$ in late October. It remained visible for 12~months until it was last seen on 8~October 1605 by Kep\-ler (Kepler 1606; Baade 1943; Stephenson \& Green 2002, chapter~5; Turatto et al.\ 2005).

\section{The Sagittarius mural at the tomb of Madin \d{S}\={a}\d{h}ib in Srinagar}

In a recent paper in this journal (Sule et al.\ 2011), the authors claim that an early 17th-century Indian glazed-tile mural (Fig.~\ref{fig:sgr_nicholls}), formerly fixed in the left spandrel of the eastern entrance arch of the shrine of the Muslim saint Sayyid Mu\d{h}ammad Ma\-dan\={\i} ($\dag$~1445, locally known as ``Madin \d{S}\={a}\-\d{h}ib'') and adjacent to the Madin \d{S}\={a}\d{h}ib mosque (completed in 1444) in the Zadibal suburb of Srinagar (Kashmir Valley, India), preserves a unique Indian pictorial record of the stellar outburst of 1604.
 
The mural was first described in 
situ\footnote{According to Mustafizur Rahman (1989) and Bandey (2012) the Sagittarius mural was removed in the 1980s to prevent further degradation by weather and vandalism. Currently, it is on display in the Central Asian Museum (University of Kashmir, Srinagar). Other decorative tiles from the same site are preserved in the Shri Pratap Singh Museum (Srinagar) and in the Victoria and Albert Museum (London).}
by the British architect and civil servant William Henry Nicholls (1865-1949), who visited the site in 1905. In his report, he 
wrote (Nicholls 1909):

\begin{figure*}[ht]
\begin{center}
\frame{\includegraphics[angle=0,width=0.99\textwidth]{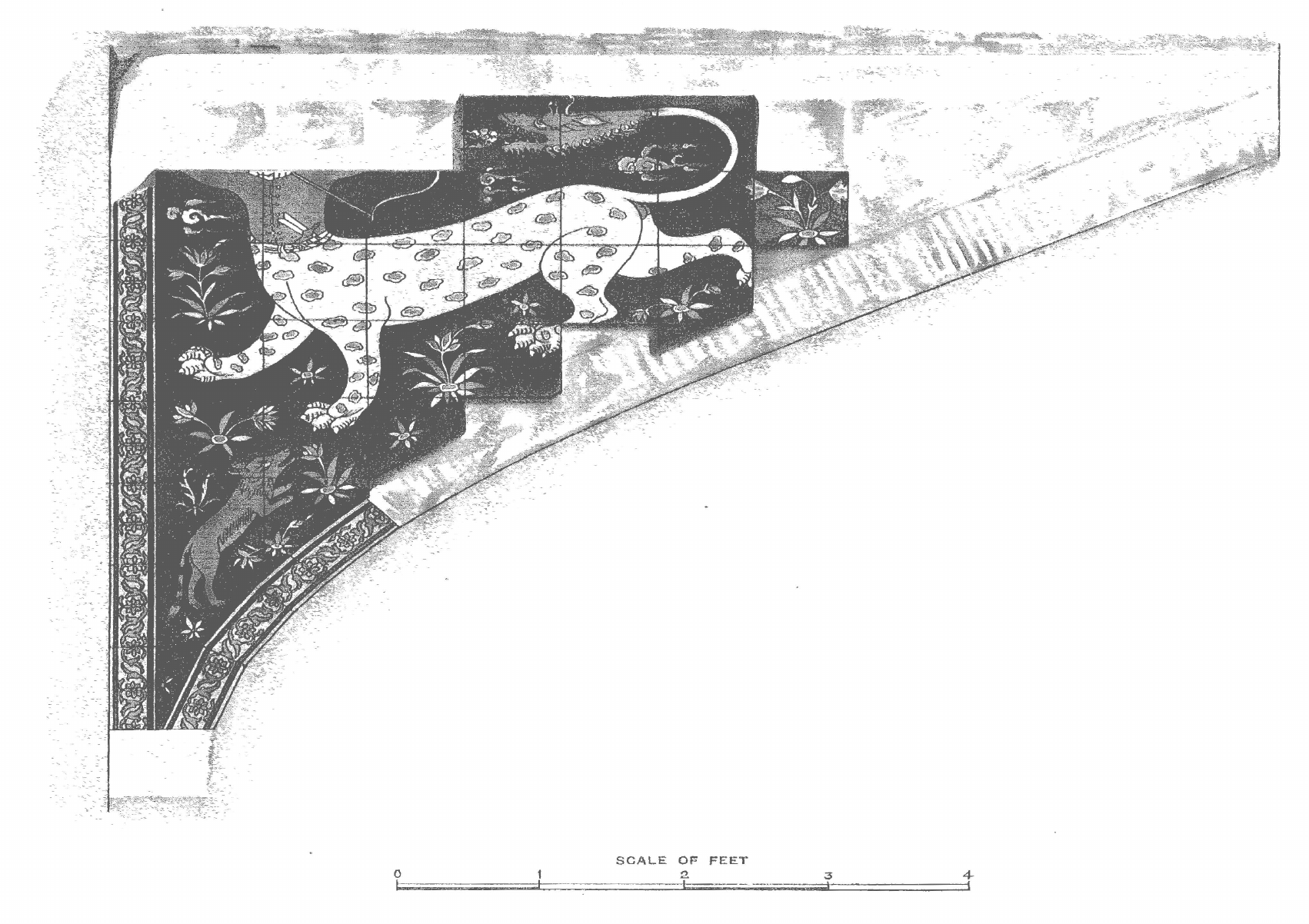}}
\caption{Glazed-tile mural in the left spandrel of the eastern entrance arch of the tomb of Madin \d{S}\={a}\d{h}ib in Zadibal (Srinagar, India) as drawn in 1905 by W.H.~Nicholls. Plate~LVII from Nicholls (1909).\label{fig:sgr_nicholls}} 
\end{center}
\end{figure*}

\begin{quote}
\begin{small}
``The tilework at the tomb of Madan\={\i}, near But Kadal in Srinagar, is made in squares with various brilliant colours in contact with each other on the same piece of tile. But its great interest lies in the subject which is represented in the southern half of the spandrel of the great archway in the east fa\c{c}ade. It is hardly necessary to remind the reader that animal life was rarely represented in any form of decoration during Muhammadan rule in India. Akbar did not object to statues of horses and elephants; Jah\={a}ng\={\i}r allowed birds and butterflies to be carved, and \underline{Sh}ah Jah\={a}n also had elephants set up, and at Lahore Fort he indulged in a panelled frieze representing elephant fights, and other subjects, all in tilework. Aurangzeb was a bigot, who not only would have none of animal life in any form on his buildings, but took a delight in smashing any instances of it which came to his notice whether on Hindu or Muhammadan buildings. It is fortunate indeed that he never chan\-ced to see the tomb of Madan\={\i} when he was in Srinagar. His indignation would surely have been roused at finding, on the tomb of a Muhammadan saint, a representation of a beast with the body of a leopard, changing at the neck into the trunk of a human being, shooting apparently with a bow and arrow at its own tail, while a fox is quietly looking on among flowers and cloud-forms. These peculiar cloud-forms are common in Chinese and Persian art, and were frequently used by the Mu\underline{gh}als---by Akbar in the Turkish Sult\={a}na's house at Fat\d{h}pur-S\={\i}kr\={\i}, Jah\={a}ng\={\i}r at Sikandarah, and \underline{Sh}ah Jah\={a}n in the D\={\i}w\={a}n-i-\underline{kh}\={a}\textsubumlaut{s}\textsubumlaut{s} at Delhi, to mention only a few instances. The principal beast in the picture is about four feet long, and is striking quite an heraldic attitude. The chest, shoulders, and head of the human being are unfortunately missing. The tail ends in a kind of dragon's head. As for the colours, the background is blue, the trunk of the man is red, the leopard's body is yellow with light green spots, the dragon's head and the fox are reddish brown, and the flowers are of various colours. It is most probable that if this beast can be run to earth, and similar pictures found in the art of other countries, some light will be thrown upon the influences bearing upon the architecture of Kashmir during a period about which little is at present known.''
\end{small}
\end{quote}

Nicholls assumed that the tilework on Madan\={\i}'s tomb dated from \textit{c.}~1445 but in his introduction to Nicholls' report, John Hubert Marshall (1876-1958), superintendent of the Archaeological Survey of India, referred to a Persian text at the site indicating that the present entrance was added during Sh\={a}h Jah\={a}n's reign (1628 to 1658).

Although Nicholls was unaware of the celestial origin of the mural, later researchers (Bandey 1994; Sule et al.\ 2011) correctly identified it as a representation of the zodiacal constellation Sagittarius which is traditionally depicted as a centaur, a hybrid creature whose upper part is human and whose lower part is equine.

Sule et al.\ (2011) claim that the dragon-headed tail is such an unusual addition to the figure of Sagittarius in Indian art, that it must depict some extraordinary and temporary feature in the night sky and thus can only refer to the supernova of 1604 and the planets Mars, Jupiter and Saturn which happened to be nearby when the new star first appeared in the neighbouring constellation of Ophiuchus. 

However, this claim is based on an analysis of early-modern constellation images in which only the astronomical sources were studied while other equally relevant astrological sources were neglected. 

\section{Depictions of Sagittarius in Islamic art}

In general, depictions of the zodiacal constellation Sagittarius in Islamic art and manuscripts can be divided into two distinct families: an \textit{astronomical group} closely following the traditional (Hellenistic) constellation imagery and an \textit{astrological group} in which additional features (referring to the astrological nature of the sign) were added to the constellation figure.

\begin{figure}
\begin{center}
\frame{\includegraphics[angle=0,width=0.48\textwidth]{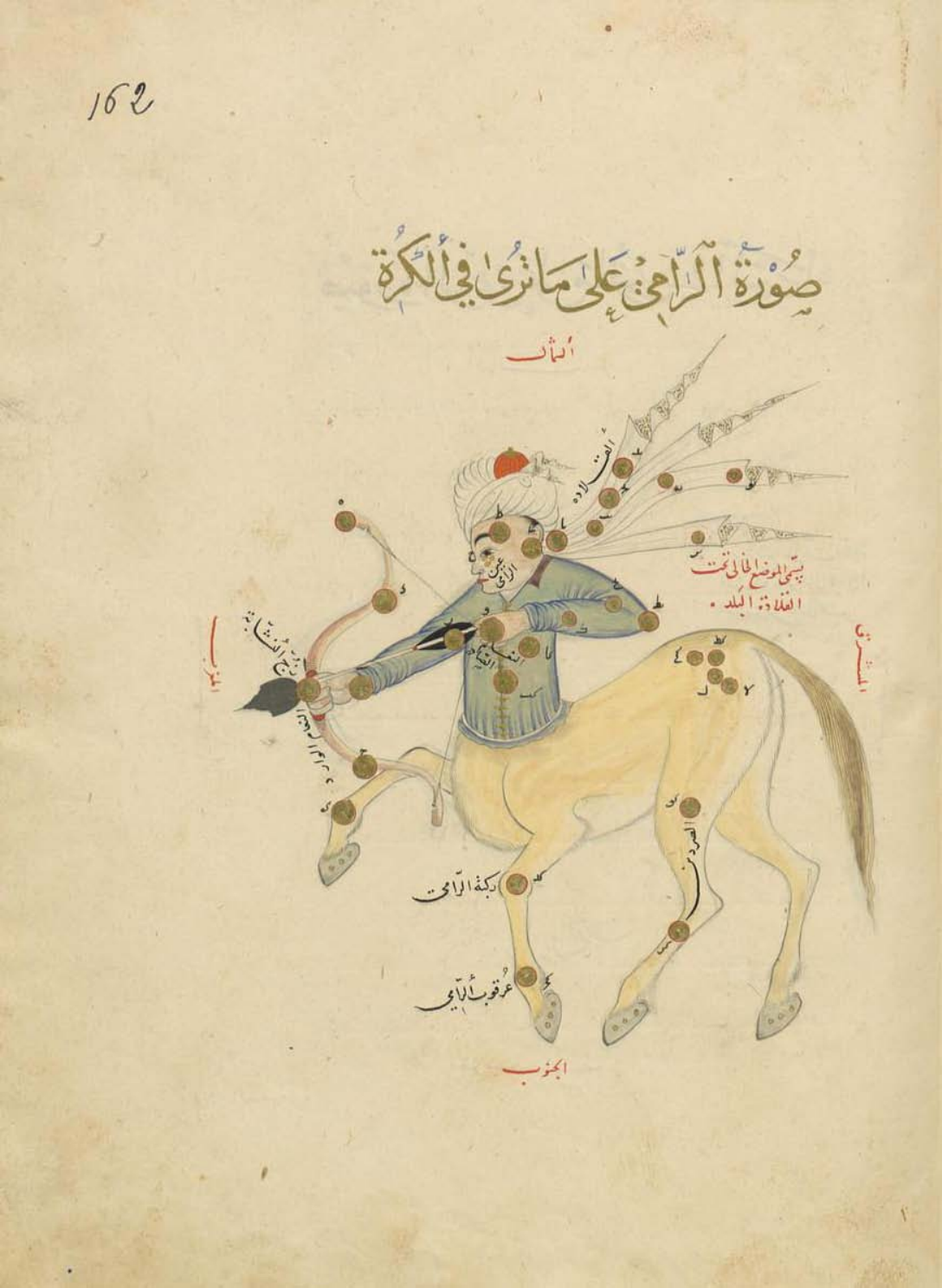}}
\caption{The constellation of Sagittarius (as viewed in the sky) in a manuscript of al-\d{S}\={u}f\={\i}'s \textit{Kit\={a}b \d{s}uwar al-kaw\={a}kib al-th\={a}bitah}, copied in Samarkand between 1430 and 1440. Paris, Biblioth\`{e}que Nationale de France (ms.\ Arabe~5036, fol.~162$^{\rm r}$).\label{fig:sgr_alsufi}} 
\end{center}
\end{figure}

\begin{table*}
\begin{center}
\caption{The zodiacal signs in Islamic astrology\label{tab:signs}}
\begin{small}
\begin{tabular}{llllll}
\hline
\multicolumn{1}{c}{Zodiacal sign} &
\multicolumn{1}{c}{Sign} &
\multicolumn{3}{c}{Decan rulers} &
\multicolumn{1}{c}{Exaltation}\\
\multicolumn{1}{c}{} &
\multicolumn{1}{c}{ruler} &
\multicolumn{1}{c}{first} &
\multicolumn{1}{c}{second} &
\multicolumn{1}{c}{third} &
\multicolumn{1}{c}{}\\
\hline\hline
Aries (\textit{al-\d{H}amal})            & Mars    & Mars    & Sun     & Venus   & Sun (19\degr)\\
Taurus (\textit{al-Thawr})               & Venus   & Mercury & Moon    & Saturn  & Moon (3\degr)\\
Gemini (\textit{al-Tawa'm\={a}n})        & Mercury & Jupiter & Mars    & Sun     & Lunar Node (ascending) (3\degr)\\
Cancer (\textit{al-Sara\d{t}\={a}n})     & Moon    & Venus   & Mercury & Moon    & Jupiter (15\degr)\\
Leo (\textit{al-Asad})                   & Sun     & Saturn  & Jupiter & Mars    & \\
Virgo (\textit{al-`Adhr\={a}'})          & Mercury & Sun     & Venus   & Mercury & Mercury (15\degr)\\
Libra (\textit{al-M\={\i}z\={a}n})       & Venus   & Moon    & Saturn  & Jupiter & Saturn (21\degr)\\
Scorpius (\textit{al-`Aqrab})            & Mars    & Mars    & Sun     & Venus   & \\
Sagittarius (\textit{al-R\={a}m\={\i}})  & Jupiter & Mercury & Moon    & Saturn  & Lunar Node (descending) (3\degr)\\
Capricornus (\textit{al-Jad\={\i}})      & Saturn  & Jupiter & Mars    & Sun     & Mars (28\degr)\\
Aquarius (\textit{S\={a}kib al-M\={a}'}) & Saturn  & Venus   & Mercury & Moon    & \\
Pisces (\textit{al-Samakat\={a}n})       & Jupiter & Saturn  & Jupiter & Mars    & Venus (27\degr)\\
\hline
\end{tabular}
\end{small}
\end{center}
\end{table*}

\subsection{Constellation iconography in Islamic astronomical manuscripts}

The \textit{astronomical tradition} in Islamic constellation imagery stems from the \textit{Kit\={a}b \d{s}u\-war al-kaw\={a}kib al-th\={a}bitah} (``Book of the Images of the Fixed Stars'') of the Persian astronomer Ab\={u} al-\d{H}usayn `Abd al-Ra\d{h}m\={a}n ibn `Umar al-\d{S}\={u}f\={\i} (903-986), who was known in the West as Azophi. This influential uranographical treatise, completed in Sh\={\i}r\={a}z (Persia) around 964, contains an Arabic edition of the star catalogue of Claudius Ptolemy (books~VII \& VIII in the \textit{Almagest}) with al-\d{S}\={u}f\={\i}'s own stellar magnitude estimates and additional information on Arabic star lore drawn from early Islamic sources (Schjellerup 1874; Kunitzsch 1986).

Already in the earliest-known manuscript copies, this work routinely includes two mirrored depictions of each of the 48 Ptolemaic constellations: one as it is seen in the sky (`internal' view) and one as it is seen on a celestial globe (`external' view). As al-\d{S}\={u}f\={\i}'s star catalogue is a copy of Ptolemy's star catalogue, precessed to the epoch 1276 of Alexander [= 1~October 964], the constellation figures closely follow the Greek 
tradition (Upton 1932/33; Wellesz 1959; Wellesz 1965; Carey 2001) and never add a dragon's head to the centaur's tail (Fig.~\ref{fig:sgr_alsufi}). This is also true for the depiction of Sagittarius on Islamic celestial globes, which in general closely follow the iconography of al-\d{S}\={u}f\={\i}'s star atlas (Savage-Smith 1985).

\subsection{Constellation iconography in Islamic astrological manuscripts}

The \textit{astrological tradition} in Islamic constellation imagery is derived from early-medieval Islamic astrology, which was largely based on Hellenistic and late-classical astrological (Bouch\'{e}-Leclerq 1899; Boll et al.\ 1931; Neugebauer \& van Hoesen 1959; Gundel \& Gundel 1966) and also from Late-Babylonian traditions which were still preserved in Harran (ancient Carrhae in southern Anatolia), an ancient site of star worship (Chwolsohn 1856; Green 1992; Pingree 2002) and place of origin of some of the most skilled translators of scientific works in Baghdad during the Abbasid period. 

The most influential author of Islamic astrological lore was the Persian astrologer Ab\={u} Ma`shar Ja`far ibn Mu\d{h}am\-mad ibn `Umar al-Balkh\={\i} (787-886), who was known in the West as Albumasar. Of his many works on astrology, some of which were also translated in Latin from the late 12th century onwards, the best known are his \textit{Kit\={a}b al-mudkhal al-kab\={\i}r ila `ilm a\d{h}k\={a}m al-nuj\={u}m} (``Great Introduction to Astrology''), his \textit{Kit\={a}b al-mudkhal al-\d{s}agh\={\i}r} (``Abbreviation of the Introduction [to Astrology]'') and his \textit{Kit\={a}b al-maw\={a}-lid} (``Book of Nativities''). Detailed expositions of Islamic astrology can be found in modern editions of Ab\-{u} Ma'shar's 
works (Burnett et al.\ 1994; Lemay 1995/96) or of those of later Islamic astrologers (Ramsey Wright 1934; Burnett et al.\ 2004).

In order to understand the iconography of Islamic astrological manuscripts it will suffice to list in Table~\ref{tab:signs} the zodiacal signs and their planetary rulers, their tripartite divisions (the decans) and associated planets (decan rulers) and the locations of the so-called exaltations. In both the sign and the decan rulers it is easy to recognize the traditional planetary order (Saturn, Jupiter, Mars, Sun, Venus, Mercury, Moon), based on their assumed distances from the centre of the cosmos (the Earth). 

Following an ancient tradition, certain points on the zodiac known as the `exaltations' (or \textit{hypsomata}) of the planets were of special importance. Already mentioned in Assyrian astrological tablets dating from the 7th century BC as the `secret houses' (\textit{b\={\i}t ni\d{s}irti}) of the planets, these were places where a planet was believed to indicate good 
fortune (Rochberg 1998). Specific longitudes within a sign are already cited in classical texts dating from the 1st century 
AD\footnote{Pliny the Elder, \textit{Naturalis historiae} II~13 [65].}
and these were also adopted by Islamic astrologers who added values for the lunar nodes. 

In Islamic astrological sources the ascending and descending nodes of the lunar orbit were also regarded as planets (Hartner 1938; Hartner 1965; Hartner 1978; Kennedy 1956; Neugebauer 1957; Azarpay \& Kilmer 1978) whose exaltations were located in Gemini and Sagittarius. They were known as \textit{ra's al-jawzahar} (``the dragon's head'') and \textit{dhanab al-jawzahar} (``the dragon's tail''), from which the European Latin translations \textit{caput draconis} and \textit{cauda draconis} were derived.

\begin{figure}
\begin{center}
\frame{\includegraphics[angle=0,width=0.48\textwidth]{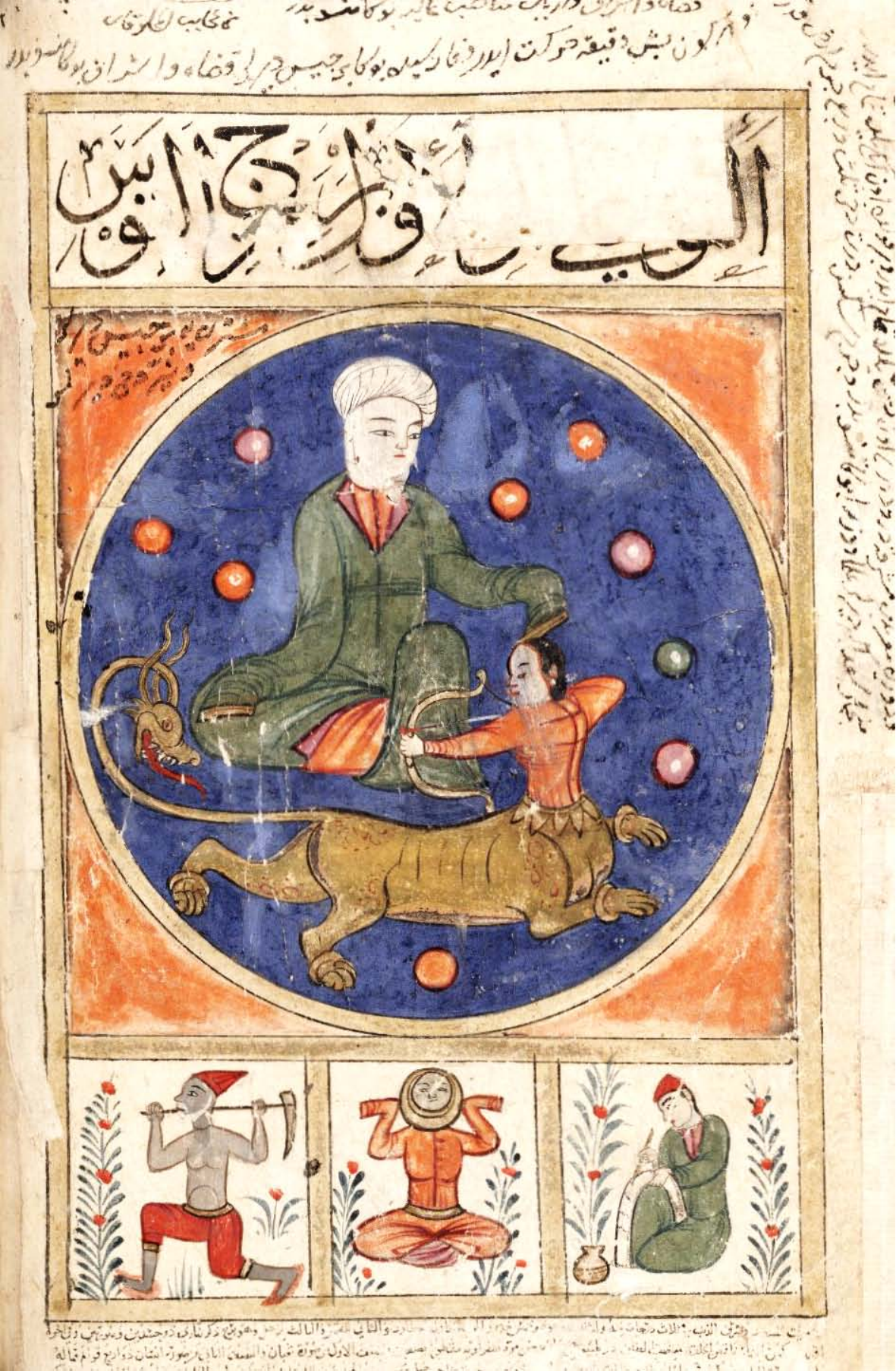}}
\caption{The zodiacal sign Sagittarius in the \textit{Kit\={a}b al-bulh\={a}n} (``Book of Wonders''), an Islamic divinatory compendium dated to the late 14th or early 15th century, with its sign ruler Jupiter. Below (from right to left) are the decan rulers Mercury, the Moon and Saturn. Oxford, Bodleian Library (ms.\ Or.~133, fol.~17$^{\rm v}$).\label{fig:sgr_abumashar1}} 
\end{center}
\end{figure}

\begin{figure}
\begin{center}
\frame{\includegraphics[angle=0,width=0.48\textwidth]{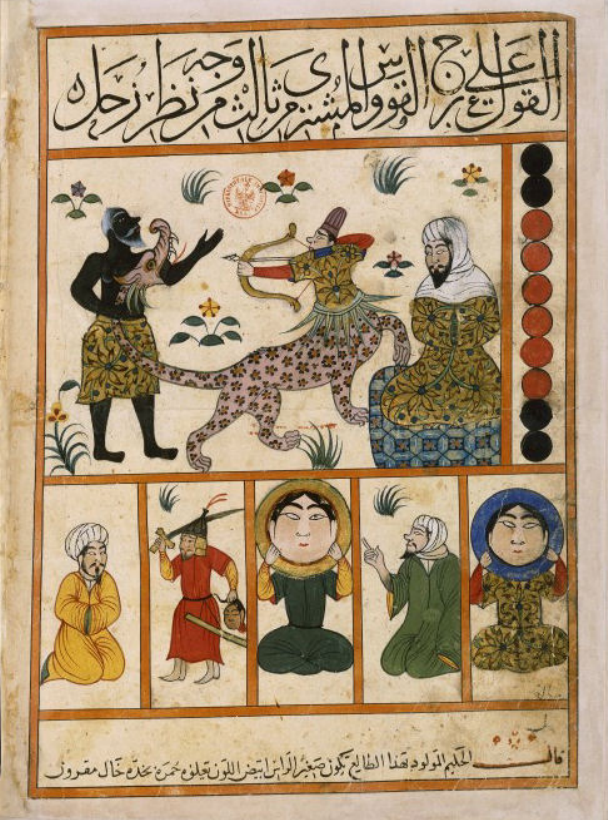}}
\caption{The third decan of the zodiacal sign Sagittarius in a manuscript (dated AH~700 [1300/01~CE]) of Ab\={u} Ma`shar's \textit{Kit\={a}b al-maw\={a}lid} (``Book of Nativities''). Jupiter, the ruler of the sign, is seated at the right and Saturn, the decan ruler, stands at the left. Depicted below (from right to left) are the Moon, Mercury, the Sun, Mars and Jupiter, a sub-set of the planetary `terms' (\textit{\d{h}ud\={u}d}) for this decan attributed to As\d{t}ar\={a}\d{t}\={u}s (Erasistratos). The coloured circles denote individual degrees in this decan, the significance of which is not well understood. Paris, Biblioth\`{e}que Nationale de France (ms.\ Arabe~2583, fol.~27$^{\rm v}$).\label{fig:sgr_abumashar2}} 
\end{center}
\end{figure}

Figs.~\ref{fig:sgr_abumashar1} and \ref{fig:sgr_abumashar2} offer typical examples of how the zodiacal sign of Sagittarius was depicted in Islamic astrological manuscripts. A notable feature of these representations of Sagittarius is the dragon-headed tail and the spotted (or striped) coat of the animal's usually feline (rather than an equine) body, which is found in nearly all known examples. Representations of the diametrically opposite zodiacal sign (Gemini) also usually feature a dragon-like or demonic
head.\footnote{Other peculiarities in the depiction of the zodiacal constellations often found in Islamic astrological manuscripts include: Virgo depicted as a young man sitting in a wheat field, Libra upheld by a young woman, Scorpius as two scorpions encircling each other and Aquarius as a man drawing water from a well.}

The planets in these manuscripts also have a very distinctive iconography, which bear little relation to the Greek-Roman representations found in classical art and in early-medieval Greek and Latin manuscripts, but rather seem to refer to Late-Babylonian descriptions of the 
planets (Saxl 1912; Ruska 1919; Carboni 1997; but see also Blume 2000, Exkurs~1; Caiozzo 2003). In Islamic astrological texts both the Moon (\textit{al-Qamar}) and the Sun (\textit{al-Shams}) are always recognizable by their circular faces, Mercury (\textit{al-`U\d{t}\={a}rid}) is usually depicted as a seated scribe, Venus (\textit{al-Zuhara}) always plays a musical instrument (lute or harp), Mars (\textit{al-Mirr\={\i}kh}) usually wields a sword and carries a decapitated head, Jupiter (\textit{al-Mushtar\={\i}}) is often depicted as a person in contemplation (or observing with an astrolabe) while Saturn (\textit{al-Zu\d{h}al}) is usually dark-skinned and carrying an axe. 

\begin{figure}[t]
\begin{center}
\frame{\includegraphics[angle=0,width=0.48\textwidth]{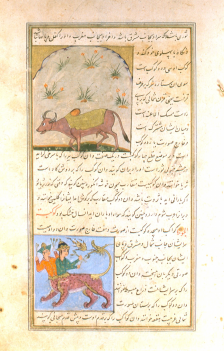}}
\caption{The zodiacal signs Taurus and Gemini in a Persian manuscript (dated AH~1010 [1601/02~CE]) of al-Qazw\={\i}n\={\i}'s \textit{Kit\={a}b `aj\={a}ib al-makhl\={u}q\={a}t}. Leiden, University Library (ms.\ Or.~8907, fol.~21$^{\rm r}$).\label{fig:gem_qazwini}} 
\end{center}
\end{figure}

\begin{figure}[t]
\begin{center}
\frame{\includegraphics[angle=0,width=0.48\textwidth]{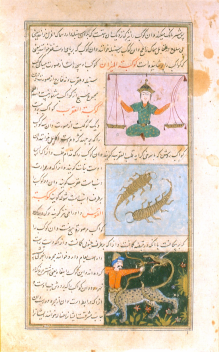}}
\caption{The zodiacal signs Libra, Scorpius and Sagittarius in a Persian manuscript of al-Qazw\={\i}n\={\i}'s \textit{Kit\={a}b `aj\={a}ib al-makhl\={u}q\={a}t}. Leiden, University Library (ms.\ Or.~8907, fol.~22$^{\rm r}$).\label{fig:sgr_qazwini}} 
\end{center}
\end{figure}

Ab\={u} Ma`shar's zodiacal and planetary figures were also frequently copied in later astrological and cosmographical works such as the \textit{Kit\={a}b `aj\={a}'ib al-makhl\={u}q\={a}t wa ghar\={a}'ib al-mawj\={u}\-d\={a}t} (``Book of the Wonders of Creation and the Rarities of the World'') by Ab\={u} Ya\d{h}y\={a} Zakariyy\={a} ibn Mu\-\d{h}ammad ibn Ma\d{h}m\={u}d al-Qazw\={\i}n\={\i} (1203-1283), a Persian imam, jurist and cosmographer (Eth\'{e} 1868). This work gave a description of the cosmos, the earth and its inhabitants which, judging from the large number of manuscript copies that are still extant, enjoyed great popularity in the Islamic world (Carboni 1992; Caiozzo 2003). Figs.~\ref{fig:gem_qazwini} \& \ref{fig:sgr_qazwini} give some typical examples. 

\section{Conclusion}

From the above-discussed depictions of Sagittarius in Islamic astrology, it is evident that the Sagittarius mural at Srinagar is derived from an Islamic astrological source and therefore cannot be used as evidence for an Indian sighting of SN1604. 

Although the completeness with which Far-Eastern astronomers re\-corded various transient celestial phenomena (eclipses, co\-mets, new stars, meteors, sunspots and aurorae) is unrivaled, Islamic astronomers and historians also made numerous observations of unusual celestial events. So we have fairly comprehensive records of lunar and solar eclipses (Ginzel 1887; Said et al.\ 1989; Stephenson \& Said 1989; Stephenson \& Said 1991ab; Said \& Stephenson 1996; Said \& Stephenson 1997; Stephenson \& Said 1997), comets and meteors (Modi 1910ab; Khan 1946; Khan 1948; Kennedy 1956; Rada \& Stephenson 1992; Kidger 1993; Ha\-segawa 1996; Cook 1999; Rada 1999/2000) and sunspots (Vaquero \& Gallego 2002ab).

However, Islamic records of `new stars' are indeed rare and are limited to reports of the outbursts of SN1006 (Goldstein 1965; Porter 1974) and SN1054 (Brecher et al.\ 1978).

In general, Hindu astronomers do not appear to have been greatly interested in uranography (Kaye 1924). The Persian polymath Ab\={u}'l-Ray\d{h}\={a}n Mu\d{h}ammad ibn A\d{h}mad al-B\={\i}r\={u}n\={\i} (973-1048), who visited the western and central regions of India in the 1030s and admired Hindu astronomers for their knowledge of the planetary motions, nevertheless noted in his \textit{Kit\={a}b f\={\i} ta\d{h}r\={\i}r m\={a} li'l-Hind min maq\={a}la maqb\={u}\-la f\={\i}'l-`aql aw mardh\={u}la} (``The Book Containing Explanation of Doctrines of Indians, Both Acceptable by Reason or Rejectable''):

\begin{quote}
\begin{small}
``The Hindus are very little informed regarding the fixed stars. I never came across any one of them who knew the single stars of the lunar stations from eyesight, and was able to point them out to me with his fingers.'' (Sachau 1888, vol.~2, 83)
\end{small}
\end{quote}

It is also significant that the \textit{Akbarn\={a}ma}, the voluminous official history of the reign of the third Mughal emperor Akbar the Great (reigned 1556-1605), mentions the great comet of 1577 in detail (Beveridge 1907/39, vol.~3, 311), and some earlier comets in AH~662 (1263/64~CE), AH~803 (1400/01~CE) and AH~837 (1433/34~CE) (\textit{ibid.}, vol.~3, 314), but is silent on the new stars of 1572 and 1604, the latter occurring just a year before the emperor's death.

\acknowledgements
I am grateful to prof.\ dr.\ Aijaz Bandey (Centre for Central Asian Studies, University of Kashmir, Srinagar) for providing me with documents and recent information on the Sagittarius mural at Srinagar. I am also grateful to prof.\ dr.\ Jan Hogendijk (Department of Mathematics, Utrecht University) for comments and corrections to the draft version and to the referee for critically reading the manuscript.


\begin{thebibliography}{}
  \bibitem{} Azarpay, G., Kilmer, A.D.: 1978, J.\ American Oriental Soc.\ 98, 363
  \bibitem{} Baade, W.: 1943, ApJ 97, 119
  \bibitem{} Bandey, A.A.: 1994, J.\ Central Asian Studies 5, 81
  \bibitem{} Bandey, A.A.: 2012, pers.\ comm.
  \bibitem{} Beveridge, H.: 1907/39, \textit{Akbar-N\^{a}ma: A History of the Reign of Akbar including an Account of his Predecessors by Abu `l-Fazl},
             Asiatic Society, Calcutta, 3~vols
  \bibitem{} Blume, D.: 2000, \textit{Regenten des Himmels: Astrologische Bilder in Mittelalter und Renaissance}, Akademie Verlag, Berlin
  \bibitem{} Boll, Fr., Bezold, C., Gundel, W.: 1931, \textit{Sternglaube und Sterndeutung: Die Geschichte und das Wesen der Astrologie},
             B.G.~Teubner, Leipzig/Berlin
  \bibitem{} Bouch\'{e}-Leclercq, A.: 1899, \textit{l'Astrologie grecque}, Ernest Leroux, Paris
  \bibitem{} Brecher, K., Lieber, E., Lieber, A.E.: 1978, Nature 273, 728
  \bibitem{} Burnett, Ch., Yamamoto, K., Yano, M.: 1994, \textit{Ab\={u} Ma`\v{s}ar: The Abbreviation of the Introduction to Astrology, together with
             the Medieval Latin Translation of Adelard of Bath}, E.J.~Brill, Leiden [etc.]
  \bibitem{} Burnett, Ch., Yamamoto, K., Yano, M.: 2004, \textit{Al-Qab\={\i}\d{s}\={\i} (Alcabitius): The Introduction to Astrology. Editions of the
             Arabic and Latin Texts and an English Translation}, Warburg Institute/Nino Aragno Editore, London/Turin
  \bibitem{} Caiozzo, A.: 2003, \textit{Images du ciel d'Orient au Moyen \^{A}ge: Un histoire du zodiaque et de ses repr\'{e}sentations dans les
             manuscrits du Proche-Orient musulman}, Presse de l'Universit\'{e} de Paris-Sorbonne, Paris
  \bibitem{} Carboni, S.: 1992, \textit{The Wonders of Creation and the Singularities of Ilkhanid Painting: A Study of the London Qazw\={\i}n\={\i}
             (British Library ms.\ Or.~14140)}, School of Oriental and African Studies, London
  \bibitem{} Carboni, S.: 1997, \textit{Following the Stars: Images of the Zodiac in Islamic Art}, Metropolitan Museum of Art, New York
  \bibitem{} Carey, M.C.: 2001, \textit{Painting the Stars in a Century of Change: A Thirteenth-Century Copy of al-\d{S}\={u}f\={\i}'s Treatise on
             the Fixed Stars (British Library Or.~5323)}, School of Oriental and African Studies, London
  \bibitem{} Chwolsohn, D.: 1856, \textit{Die Ssabier und der Ssabismus}, Buchdruckerei der Kaiserlichen Akademie der Wissenschaften, St.~Petersburg,
             2~vols
  \bibitem{} Cook, D.: 1999, J.\ Hist.\ Astron.\ 30, 131
  \bibitem{} Eth\'{e}, H.: 1868, \textit{Zakarija ben Muhammed ben Mahm\^{u}d al-Kaz\-w\^{\i}ni's Kosmographie: Erster Halbband. Die Wunder der
             Sch\"{o}p\-fung}, Fues's Verlag, Leipzig
  \bibitem{} Ginzel, Fr.K.: 1887, Sitzungsberichte der K\"{o}niglich Preu{\ss}ischen Akademie der Wissenschaften zu Berlin, Theil~II, 709
  \bibitem{} Goldstein, B.R.: 1965, Astron.\ J.\ 70, 105
  \bibitem{} Green, T.M.: 1992, \textit{The City of the Moon God: Religious Traditions of Harran}, E.J.~Brill, Leiden
  \bibitem{} Gundel, W., Gundel, H.G.: 1966, \textit{Astrologumena: Die astrologi\-sche Literatur in der Antike und ihre Geschichte}, Franz Steiner
             Verlag, Wiesbaden
  \bibitem{} Hartner, W.: 1938, Ars Islamica 5, 112
  \bibitem{} Hartner, W.: 1965, in \textit{The Encyclopaedia of Islam: New Edition}, E.J.~Brill, Leiden, vol.~2, 502 [``al-\underline{Dj}awzahar'']
  \bibitem{} Hartner, W.: 1978, in \textit{The Encyclopaedia of Islam: New Edition}, E.J.~Brill, Leiden, vol.~4, 809 [``al-Kayd'']
  \bibitem{} Hasegawa, I.: 1996, Quart.\ J.\ Roy.\ Astron.\ Soc.\ 37, 75
  \bibitem{} Kaye, G.R.: 1924, \textit{Hindu Astronomy}, Government of India Central Publication Branch, Calcutta, 24
  \bibitem{} Kennedy, E.S.: 1956, J.\ Near East.\ Studies 16, 44
  \bibitem{} Kepler, J.: 1606, \textit{De Stella Nova in pede Serpentarii, et qui sub ejus exortum de novo iniit, Trigono igneo}, Paul Sessi, Prague                 [reprinted in M.~Caspar (ed.): 1938, \textit{Johannes Kepler Gesammelte Werke: Band~I, Mysterium Cosmographicum / De Stella Nova},
             C.H.~Beck'sche Verlagsbuchhandlung, Munich, 147]
  \bibitem{} Khan, M.A.R.: 1946, Islamic Culture: The Hyderabad Quart.\ Rev.\ 20, 353
  \bibitem{} Khan, M.A.R.: 1948, Islamic Culture: The Hyderabad Quart.\ Rev.\ 22, 188
  \bibitem{} Kidger, M.R.: 1993, Quart.\ J.\ Roy.\ Astron.\ Soc.\ 34, 331
  \bibitem{} Kunitzsch, P.: 1986, Z.\ Geschichte Arabisch-Islamischen Wiss.\ 3, 56
  \bibitem{} Lemay, R.: 1995/96, \textit{Ab\={u} Ma`\v{s}ar al-Bal\underline{h}\={\i} [Albumasar]: Liber Introductorii Maioris ad Scientiam Judiciorum
             Astrorum}, Istituto Universitario Orientale, Naples, 9~vols
  \bibitem{} Modi, J.J.: 1910a, Revue du Monde Musulman 10, 1 [reprinted in 1927 in \textit{Asiatic Papers: Part~III. Mostly Papers Read Before the                  Bombay Branch of the Royal Asiatic Society}, The British India Press, Bombay, 246]
  \bibitem{} Modi, J.J.: 1910b, J.\ Bombay Branch Roy.\ Asiatic Soc.\ 23, 147 [reprinted in 1917 in \textit{Asiatic Papers: Part~II. Papers Read
             Before the Bombay Branch of the Royal Asiatic Society}, The Times Press, Bombay, 69]
  \bibitem{} Mustafizur Rahman, P.I.S.: 1989, Islamic Studies 28, 171
  \bibitem{} Neugebauer, O.E.: 1957, J.\ American Oriental Soc.\ 77, 211
  \bibitem{} Neugebauer, O.E., van Hoesen, H.B.: 1959, \textit{Greek Horoscopes}, American Philosophical Society, Philadelphia
  \bibitem{} Nicholls, W.H.: 1909, \textit{Arch{\ae}ological Survey of India: Annual Report 1906-7}, Superintendent Government Printing,
             Calcutta, 42 \& 161 [reprinted in 1955 (with some additional photographs and the author's name changed to J.R.~Nichols) in
             M\={a}r\underline{g}: A Magazine of the Arts 8, nr.~2, 76]
  \bibitem{} Pingree, D.E.: 2002, Int.\ J.\ Classical Tradition 9, 8
  \bibitem{} Porter, N.A.: 1974, J.\ Hist.\ Astron.\ 5, 99
  \bibitem{} Rada, W.S.: 1999/2000, Z.\ Geschichte Arabisch-Islamischen Wiss.\ 13, 71
  \bibitem{} Rada, W.S., Stephenson, Fr.R.: 1992, Quart.\ J.\ Roy.\ Astron.\ Soc.\ 33, 5
  \bibitem{} Ramsey Wright, R.: 1934, \textit{The Book of Instruction in the Elements of the Art of Astrology by Abu'l-Ray\d{h}\={a}n Mu\d{h}ammad ibn
             A\d{h}mad al-B\={\i}r\={u}n\={\i}, Written in Ghaznah, 1029~A.D., Reproduced from Brit.\ Mus.\ MS.\ Or.~8349}, Luzac \& Co., London
  \bibitem{} Rochberg, Fr.: 1998, \textit{Babylonian Horoscopes}, American Philosophical Society, Philadelphia
  \bibitem{} Ruska, J.: 1919, \textit{Griechische Planetendarstellungen in arabischen Steinb\"{u}chern}, Carl Winter's Universit\"{a}tsbuchhandlung,
             Heidelberg
  \bibitem{} Sachau, E.C.: 1888, \textit{Alberuni's India: An Account of the Religion, Philosophy, Literature, Geography, Chronology, Astronomy,
             Customs, Laws and Astrology of India about A.D.~1030}, Tr\"{u}bner \& Co., London, 2~vols           
  \bibitem{} Said, S.S., Stephenson, Fr.R.: 1996, J.\ Hist.\ Astron.\ 27, 259
  \bibitem{} Said, S.S., Stephenson, Fr.R.: 1997, J.\ Hist.\ Astron.\ 28, 29
  \bibitem{} Said, S.S., Stephenson, Fr.R., Rada, W.S.: 1989, Bull.\ School Oriental African Studies 52, 38
  \bibitem{} Savage-Smith, E.: 1985, \textit{Islamicate Celestial Globes: Their History, Construction, and Use}, Smithsonian Institution Press,
             Washington
  \bibitem{} Saxl, Fr.: 1912, Der Islam: Zeitschrift f\"{u}r Geschichte und Kultur des Islamischen Orients 3, 151
  \bibitem{} Schjellerup, H.C.F.C.: 1874, \textit{Description des \'{e}toiles fixes compos\'{e}e au milieu du dixi\`{e}me si\`{e}cle de notre \`{e}re,
             par l'astronome persan Abd-al-Rahman Al-Sufi}, Eggers \& Cie [etc.], St.~Petersburg
  \bibitem{} Stephenson, F.R., Green, D.A.: 2002, \textit{Historical Supernovae and their Remnants}, Clarendon Press, Oxford
  \bibitem{} Stephenson, Fr.R., Said, S.S.: 1989, A\&A 215, 181
  \bibitem{} Stephenson, Fr.R., Said, S.S.: 1991a, J.\ Hist.\ Astron.\ 22, 195
  \bibitem{} Stephenson, Fr.R., Said, S.S.: 1991b, J.\ Hist.\ Astron.\ 22, 297
  \bibitem{} Stephenson, Fr.R., Said, S.S.: 1997, Bull.\ School Oriental African Studies 60, 1
  \bibitem{} Sule, A., Bandey, A.A., Vahia, M., Iqbal, N., Tabasum, M.: 2011, AN 332, 655
  \bibitem{} Turatto, M., Benetti, S., Zampieri, L., Shea, W.\ (eds.): 2005, \textit{1604-2004: Supernovae as Cosmological Lighthouses}, Astronomical
             Society of the Pacific, San Francisco
  \bibitem{} Upton, J.M.: 1932/33, Metropolitan Museum Studies 4, 179
  \bibitem{} Vaquero, J.M., Gallego, M.C.: 2002a, Astron.\ Geophys.\ 43, nr.~2, 8
  \bibitem{} Vaquero, J.M., Gallego, M.C.: 2002b, Sol.\ Phys.\ 206, 207
  \bibitem{} Wellesz, E.: 1959, Ars Orientalis 3, 1           
  \bibitem{} Wellesz, E.: 1965, \textit{An Islamic Book of Constellations}, Bodleian Library, Oxford
\end{thebibliography}
\end{document}